# The hidden magnetization in ferromagnetic material: Miamagnetism

Souri Mohamed Mimoune[1,a], Lotfi Alloui[1], Mourad Hamimid[2], Mohamed Lotfi Khene[1], Nesrine Badi[1] and Mouloud Feliachi[3]

[1]LMSE, Université de Biskra, BP 145, 07000 Biskra, Algérie
[2]Université de Bordj Bou Arréridj, 34265 Bordj Bou Arréridj, Algérie
[3]IREENA-IUT, CRTT, Boulevard de l'Université, BP 406, 44602 Saint-Nazaire Cedex, France
[a]corresponding author : s.mimoune@univ-biskra.dz

**Abstract**: This paper presents the hidden magnetization features of ferromagnetic materials: called miamagnetism. As we know, we have several forms of magnetization: the diamagnetism, the paramagnetism, the ferromagnetism etc. The main character of the diamagnetism is that its magnetic susceptibility $\chi$ is negative (from $\chi \approx -10^{-9}$ for gas and $\chi \approx -10^{-6}$ for liquid and solid to $\chi \approx -1$ for superconducting materials of type I) and it is not less than -1 unless for special materials like metamaterials at high frequencies. The miamagnetism has the character that the magnetic susceptibility can reach at low frequencies a negative value of -155 of magnitude leading to a negative permeability. We can not see it because it is hidden by the ferromagnetic character which has a high positive magnetic susceptibility. We use the discrete Fourier transform to illustrate this hidden character and the hysteresis model can be represented only by harmonics of $(2n+1)f_0$ of magnitude. This magnetization follows a Boltzmann distribution for the modulus of theses harmonics.

## 1. Introduction

In the literature, different models are used to represent the ferromagnetic character of the hysteresis. We can cite some models: the Jiles Atherton model [1], the Preisach model [2], the Hausser model [3] and other. These models are supposed to be phenomenological ones and are quite good to represent the hysteresis. This paper presents an alternative representation of the hysteresis by using Fourier series of the flux density $B$ and the magnetic field strength $H$. This representation shows a hidden character of magnetization which we call Miamagnetism. The Miamagnetism illustrates a high diamagnetic character with high negative magnetic susceptibility and magnetic permeability in low frequencies. The negative permeability is also seen in metamaterial [4] and in a composite medium interspaced split ring resonators and wires that exhibits a frequency region in the microwave regime with negative value of permeability. This paper explains also the first magnetization curve behaviour.

## 2. Purpose

When applying a sinusoidal voltage throw a conducting winding to a closed ferromagnetic material, an hysteretic behaviour occurs. The magnetomotive force acting on ferromagnetic material produces at saturated regime a non sinusoidal flux density $B$ and a distorted magnetic field $H$. Figure 1 gives the experimental result of the voltage in the secondary and the current in the primary in relative units at the frequency $f_0$ equal to 50 Hz. Figure 2 shows the voltage-current characteristic. (Note that in all that follows, the figures are in relative units).

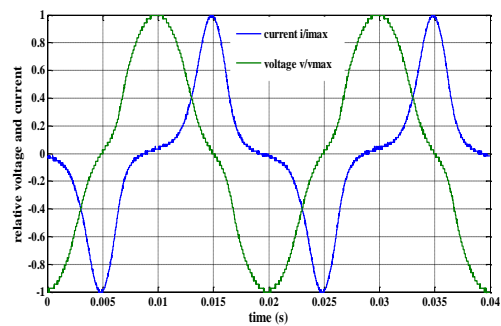

Fig. 1. Experimental voltage and current versus time.

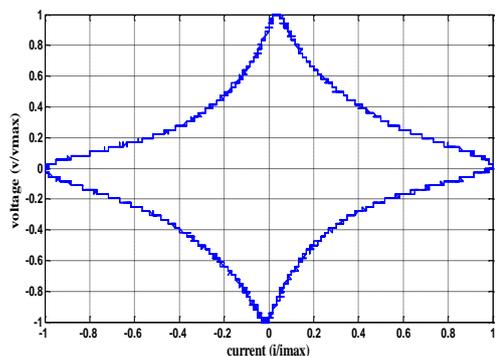

Fig. 2. Experimental current $i$ versus experimental voltage $v$.

## 3. Harmonic diagrams

We apply the discrete Fourier transform to the voltage and to the current in order to show the harmonics and then the frequencies which are important for forming these two signals. We notice that the $(2n+1)f_0$ harmonics are important for both voltage and the current (see Fig. 3).

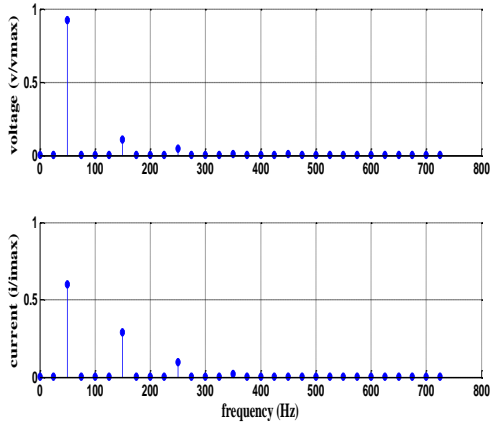

Fig. 3. Modulus diagram of the first thirteen harmonics
of the voltage *v* and the current *i*

The flux density *B* and the magnetic field *H* are derived from the voltage *v* and the current *i* respectively by the two expressions:

$$B = 1/(n_2 S) \int v \, dt$$

and $\qquad H = (n_1) i / l$

where *S* and *l* are respectively the section and the mean length of the magnetic core and $n_1$ and $n_2$ are the primary and the secondary turns respectively.

The figure 4 shows the flux density and the magnetic field versus the time. Note that the flux density is obtained by numerical integration. Figure 5 shows the hysteretic character of (*B,H*). We are not here going to redo all that has been done before and well known about hysteresis but to show other hidden aspects of this phenomenon. As we can see in figure 6 for the harmonic diagram, the flux density *B* and the magnetic field *H* revel also the importance of the $(2n+1)f_0$ harmonics. The $2nf_0$ harmonics exist but they are not fundamental for the derivation of the hysteresis.

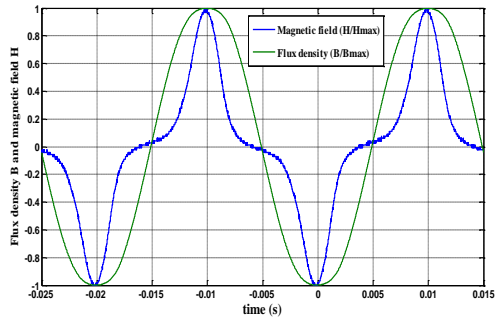

Fig.4. The flux density *B* and the magnetic field *H* versus time.

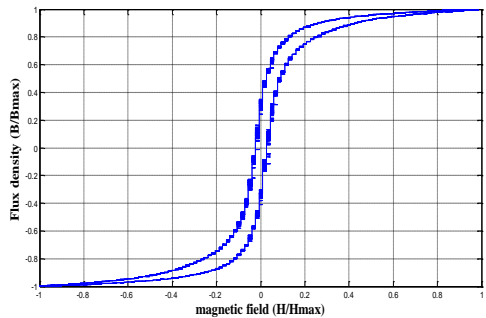

Fig.5. The hysteretic characteristic of the flux density *B* and the magnetic field *H*.

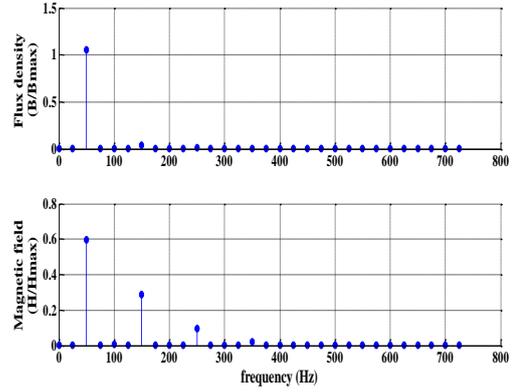

Fig.6. Modulus diagram of the flux density *B* and the magnetic field *H*.

To reconstruct these two signals we can in first approximation consider that are the sum of a finite number of harmonics having a frequency equal to (2n+1) multiple of the principal frequency $f_0$ (where n is an integer number varying from 0 to $n_{\max}$).

$$B(t) = \sum_{n=0}^{\infty} B_n(t) = \sum_{n=0}^{\infty} B_{nm} \cos(2\pi(2n+1)f_0 t + \varphi_{nb})$$

$$H(t) = \sum_{n=0}^{\infty} H_n(t) = \sum_{n=0}^{\infty} H_{nm} \cos(2\pi(2n+1)f_0 t + \varphi_{nh})$$

Figure 7 shows that to have a good approximation of the hysteresis, we need to take only the four first harmonics (*n*=0:3) and by putting the corresponding values of the modulus and phase angles of $B_n(t)$ and $H_n(t)$ we can construct the hysteresis.

$$B(t) = \sum_{n=0}^{3} B_n(t) \text{ and } H(t) = \sum_{n=0}^{3} H_n(t) \quad (1)$$

with :

$B_0 = +1.0524 \cos(2\pi(1f0)t + 91.8426)$
$B_1 = +0.0421 \cos(2\pi(3f0)t + 101,9134)$
$B_2 = +0.0105 \cos(2\pi(5f0)t - 77.3622)$
$B_3 = +0.0017 \cos(2\pi(7f0)t + 63.0582)$

$H_0 = +0.5978 \cos(2\pi(1f0)t + 96.8276)$
$H_1 = -0.2875 \cos(2\pi(3f0)t + 101.901)$
$H_2 = -0.0949 \cos(2\pi(5f0)t - 79.2867)$
$H_3 = -0.0215 \cos(2\pi(7f0)t + 82.2655)$

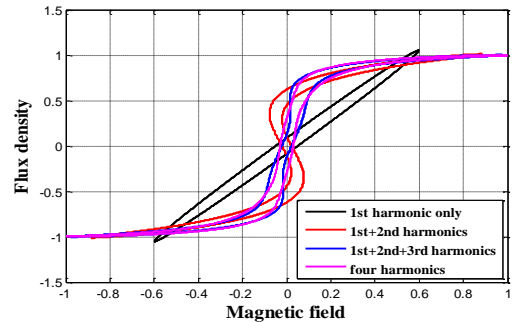

Fig.7. Reconstruction of the hysteretic characteristic by the four first harmonics.

We can see that by using only the first harmonic of the magnetic field *H* and the flux density *B* we obtain the ferromagnetic character of the hysteresis but by using the first and the second harmonics we obtain a negative slope of the hysteresis and the relative permeability can reach a hundred of magnitude. Let call the hysteresis obtained with this four harmonics the 'harmonic hysteresis'. Figure 8 shows the harmonic hysteresis compared to the experimental one.

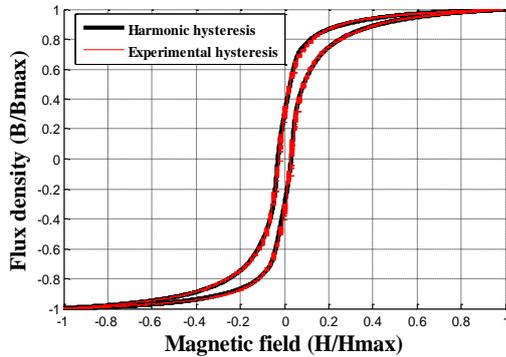

Fig. 8. harmonic hysteresis compared to the experimental one.

We can approximate the modulus of the flux density $B$ of the $(2n+1)$ frequencies ($n\leq 6$) by the following empirical expression (Fig. 9):

$$\frac{B}{B_{max}} = \frac{\left(\frac{f0}{f}\right)^2}{\exp\left(\frac{0.3(f-f0)}{f0}\right)}$$

and for the magnetic field $H$ (Fig. 10):

$$\frac{H}{H_{max}} = \frac{0.6}{\exp\left(\frac{0.5(f-f0)}{f0}\right)}$$

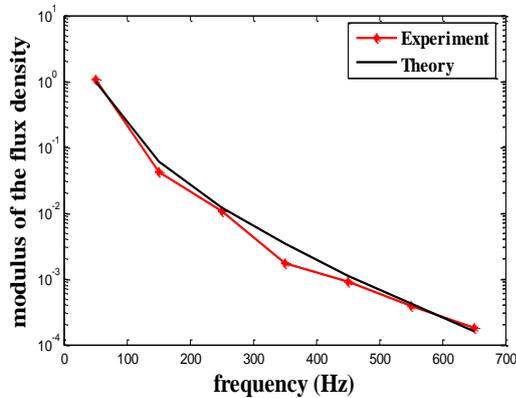

Fig. 9. Modulus of the $(2n+1)$ harmonics of the flux density.

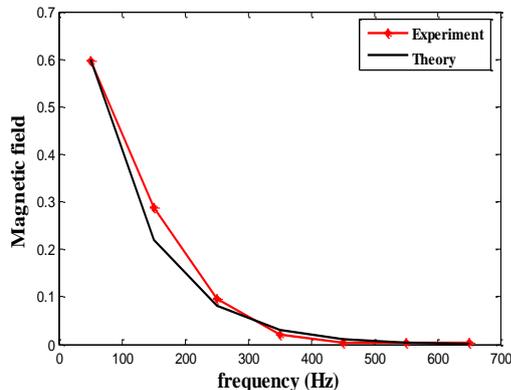

Fig. 10. Modulus of the $(2n+1)$ harmonics of the magnetic field.

This magnetization follows a Boltzmann distribution for the modulus of theses harmonics.

## 4. Miamagnetism

We will now turn to the study the harmonics of the relations (1) apart. The superposition method implies that each harmonic with a frequency different from the others acts independently on the ferromagnetic material without mutual induction with other harmonics.

If we represent the $(B_n,H_n)$ characteristics apart in figure 11, we can see that the principal harmonic characteristic $(B_0,H_0)$ has a ferromagnetic character (Fig. 11a) but for $n=1:3$ the $(B_n,H_n)$ harmonic characteristics have a very high diamagnetic character (Fig.11b). Figure 11b shows the three harmonics 1, 2 and 3 with negative slope showing increased diamagnetic appearance.

This diamagnetic character has a negative relative magnetic permeability and a high negative magnetic susceptibility ($\chi<<-1$) and can reach the average value of -155. The so-called ferromagnetic material has a ferromagnetic character with a very high positive permeability 535 which hides a very high diamagnetic character with a high negative permeability. The absolute value of this permeability is lower than the ferromagnetic permeability.

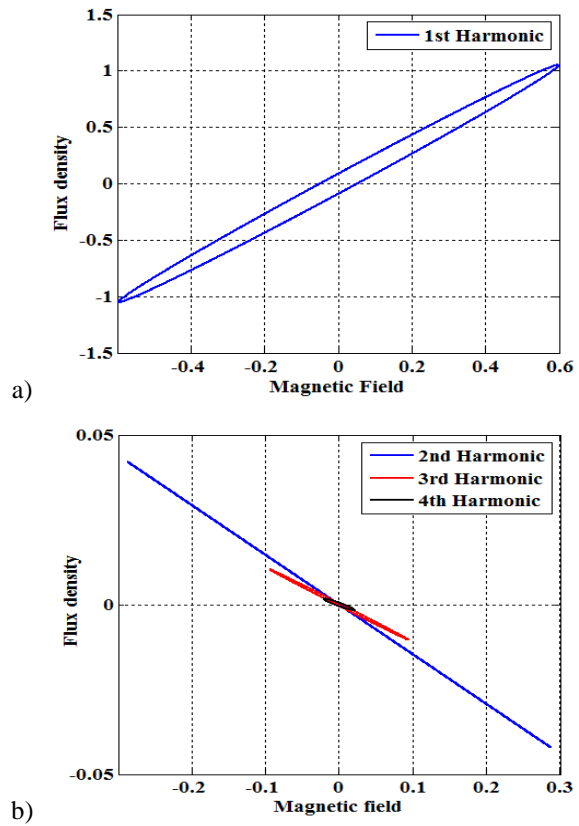

Fig.11. $(H_n,B_n)$ characteristics : a) first harmonic and b) 2nd, 3rd and 4th harmonics.

To explain this negative behavior, we suppose that the magnetic dipoles are not all oriented following the direction of the flux density **B** for all domains but a part of them are directed following the opposite direction of **B** in some domains. The delayed time is important due to their high frequencies.

## 5. Evaluation of the magnetic permeability of ferromagnetism and miamagnetism

The harmonic study allows us to determine the permeability of each harmonic. The real part of the magnetic field and flux induction signals for each harmonic makes it possible to determine the value of the real permeability whereas the imaginary part gives us information on the other losses, namely the Joule and excess losses. Table I gives the value of the real permeability for each harmonic.

For the first harmonic, we find a ferromagnetic effect which is represented by a very large positive value of the relative permeability, 535. For the other harmonics we notice an increased diamagnetic effect which is represented by a very large negative value of the average relative permeability: -154 for the second, third and fourth harmonics. To this phenomenon of diamagnetism, we attribute the name of 'Miamagnetism'. This miamagnetic behavior is hidden by the ferromagnetism which is more important. The Miamagnetism is closely related to the domains precession related to the multiple frequency of $f_0$.

Table 1. Relative permeability for each harmonic at the fundamental frequency 50 Hz.

| Harmonics | Permeability (μ) | Relative permeability ($\mu_r$) |
|---|---|---|
| 1$^{st}$ harmonic | 6.7265 10$^{-4}$ | 535.2744 |
| 2$^{nd}$ harmonic | -1.6854 10$^{-4}$ | -134.1173 |
| 3$^{rd}$ harmonic | -2.1144 10$^{-4}$ | -168.2576 |
| 4$^{th}$ harmonic | -2.0055 10$^{-4}$ | -159.5888 |

## 6. First magnetization curve

To obtain the first magnetization curve we can set the phase angle of all harmonics to 90° with respect the signs. Figure 12 gives the first magnetization curve compared to harmonic curve.

$B_0 = +1.0524 \cos(2\pi(1f0)t + 90)$
$B_1 = +0.0421 \cos(2\pi(3f0)t + 90)$
$B_2 = +0.0105 \cos(2\pi(5f0)t - 90)$
$B_3 = +0.0017 \cos(2\pi(7f0)t + 90)$

$H_0 = +0.5978 \cos(2\pi(1f0)t + 90)$
$H_1 = -0.2875 \cos(2\pi(3f0)t + 90)$
$H_2 = -0.0949 \cos(2\pi(5f0)t - 90)$
$H_3 = -0.0215 \cos(2\pi(7f0)t + 90)$

If we now suppose that the eddy current and the excess loses create a phase angle of 5° (*i.e.* -5°) we obtain an harmonic hysteresis which approach the first one (Fig. 13).

$H_0 = +0.5978 \cos(2\pi(1f0)t + 95)$
$H_1 = -0.2875 \cos(2\pi(3f0)t + 95)$
$H_2 = -0.0949 \cos(2\pi(5f0)t - 85)$
$H_3 = -0.0215 \cos(2\pi(7f0)t + 95)$

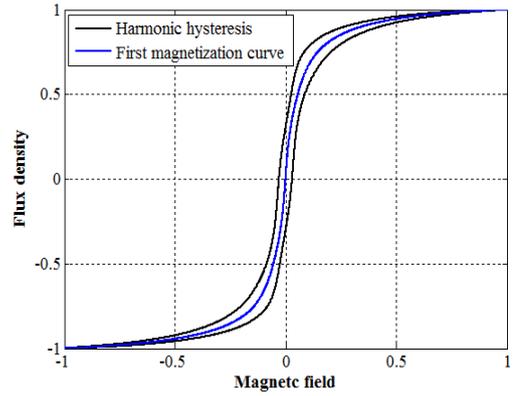
Fig. 12. harmonic hysteresis and first magnetisation curve.

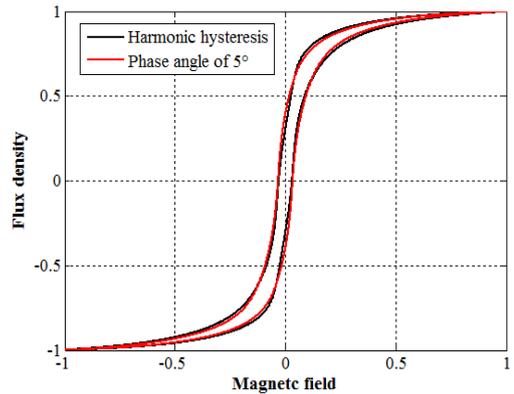
Fig. 13. harmonic hysteresis and hysteresis with phase angle of 5°.

## 7. Conclusion

The Miamagnetism has the character that the magnetic susceptibility can reach at low frequency a high negative value of -154 and greater magnitude leading to a negative permeability. We can not see it because it is hidden by the ferromagnetic character which has a higher positive magnetic susceptibility. We have used the discrete Fourier transform tff to illustrate this hidden character and the hysteresis model can only be represented by harmonics of $(2n+1)f_0$ of magnitude.

**Acknowledgements**
The authors would like to knowledge Pr. Salah Eddine Zouzou for his helpful in experimental setup.

## 8. References

[1] D. C. Jiles, D. L. Atherton, J. Appl. Phys. **55**, 2115 (1984).
[2] I. D. Mayergoyz, *Mathematical models of hysteresis*, (Springer Verlag, New York, 1991).
[3] M. Hamimid, S. M. Mimoune, M. Feliachi, *International Journal of Numerical Modelling*, John Wiley & Sons, Ltd, **30**, 6 (2017).
[4] M. S. Wartak, K. L. Tsakmakidis and O. Hess, Physics in Canada, **67**, 1 (2011).